# Fire Spread Modeling using Probabilistic Cellular Automata[*]

Rohit Ghosh[1], Jishnu Adhikary[2] and Rezki Chemlal[3]

[1] Govt. College of Engineering & Ceramic Technology
73, Abinash Chandra Banerjee Lane Kolkata-700 010, India
`rohitghosh76@gmail.com`
[2] Govt. College of Engineering & Ceramic Technology
73, Abinash Chandra Banerjee Lane Kolkata-700 010, India
`jishnuadhikary10@gmail.com`
[3] University of Béjaïa, Algeria
Laboratory of applied Mathematics, Béjaïa University, Algeria
`rchemlal@gmail.com`

**Abstract.** A cellular automaton (CA)-based modeling approach to simulate wild-fire spread, emphasizing its strengths in capturing complex fire dynamics and its integration with geographic information systems (GIS). The model introduces an enhanced CA-based methodology for wildfire prediction, emphasizing interactions between neighboring cells and incorporating major determinants of fire spread, including wind direction, wind speed, and vegetation density, while also accounting for spotting and probabilistic transitions between states in the model to mirror real-world fire behavior. This methodology is applied to case studies of the 1990 wildfire on Spetses Island, Greece, offering insights into the effects of terrain on fire spread, as well as the 2021 Evia Island wildfire in Greece, demonstrating the model's accuracy in simulating real-world wildfire scenarios.

**Keywords:** Cellular automata (CA) · Probabilistic Cellular Automata · Forest fires · Fire spread Modeling

## 1    Introduction

Forest fires have long posed significant threats, causing extensive damage to ecosystems, communities, and leading to far-reaching ecological and socio-economic consequences [1]. An alarming illustration of this destructive potential unfolded in August 2007 in Peloponnese, Greece when the nation faced its most severe forest fire catastrophe in a century, devastating approximately 2000 square kilometers of forest and agricultural land [1]. The urgency to develop and implement effective strategies to combat forest wildfires is intensifying as their frequency rises. Strategies to counteract wildfires fall into two categories: preventive measures designed to reduce the likelihood of fire ignition and immediate operational interventions in the event of a wildfire outbreak,

---

involving the efficient allocation of defensive mechanisms and the swift evacuation of vulnerable communities [4].

Developing a predictive model for wildfire expansion necessitates consideration of external environmental factors like meteorological conditions and specific terrain attributes [3]. Critical factors influencing the rate and pattern of a forest fire's progression encompass variables such as fuel type (vegetation classification), humidity, wind speed and direction, forest topography (slope and natural barriers), fuel density (vegetation thickness), and spotting—where burning materials are carried by the wind or other vectors, potentially extending the fire beyond its immediate front [2].

Constructing a comprehensive mathematical model for wildfire spread is a significant challenge, and it has received substantial attention in the literature, leading to various models. Notably, Rothermel's pioneering work [7,8] stands out. Rothermel's research established dynamic equations characterizing the maximum fire spread rate through laboratory experiments. Rothermel's equations have subsequently formed the basis for various approaches, broadly categorized into two spatial representations. The first revolves around continuous planes, assuming the fire front navigates an uninterrupted landscape in an elliptical trajectory. However, solving the associated partial differential equations can be computationally demanding. The second approach, grid-based models, offers quicker computational solutions. An alternative yet efficient approach is presented [12,13].

Within grid-based models, two subcategories emerge. The first uses the Bond–Percolation method, employing historical data for inter-grid fire spread probabilities. However, it falls short in capturing fire dynamics. The second subcategory centers on cellular automata (CA) methodology, offering a more dynamic approach with a discrete grid divided into cells governed by evolving variables and interactions with neighbouring cells [4]. This, combined with seamless integration with geographic information systems (GIS) and other data sources, makes CA an attractive candidate for modeling intricate wildfire behaviours [9,10].

This study introduces an enhanced CA-based methodology for predicting wildfire spread, emphasizing interactions between neighbouring cells and incorporating major determinants of fire spread, including spotting. The model is applied to simulate a wildfire's progression in Spetses in 1990, which offers insights into the effects of terrain slope on fire spread. Additionally, the model's parameters are fine-tuned through a nonlinear optimization approach, comparing the model's outputs with actual fire front data [4]. This work underscores the potential of CA-based models in effectively predicting wildfire dynamics and informing wildfire management strategies, especially in light of recent global fire incidents.

## 2    CA-based methodology for wildfire prediction

The predictive methodology for wildfire spread presented here employs a two-dimensional grid that subdivides the forest area into a multitude of cells. Each cell represents a small land patch, chosen to be square in shape, providing eight possible directions for fire propagation (refer to Fig. 2). Certain researchers have used [11] pentagonal or hexagonal cells, as they provide a more precise representation of spatial fire behaviour. However, it's important to acknowledge that such a choice significantly escalates computational complexity. In this paper, we favour the use of a square grid for its ability to simplify calculations.

## 2.1 Grid Definition

In the simulation, each cell on the 2D grid is associated with a specific state, and these states are represented by various categories. Here are the details of each state and its corresponding category:

- TREE: Contains unignited forest fuel.
- BURNING: Actively burning forest fuel.
- BURNING DURATION: Once the tree cell has been ignited by the fire, the cell transitions to this category, and the duration can be controlled probabilistically.
- EMPTY: Urban and non-vegetated areas with no forest fuel.
- WATER: Non-combustible water bodies.
- CITY: Urban zones or settlements within the simulation. These regions often have non-combustible properties and serve as obstacles to the fire's propagation.

The state of each cell is expressed as an element in a matrix denoted as the state matrix, specifically utilizing the Moore matrix to represent the directions of fire spread. An illustration of an area comprising 25 cells encoded in matrix form.

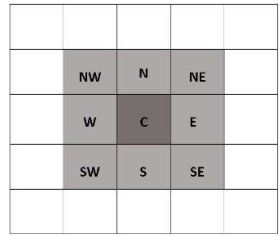

**Fig. 1.** Possible directions of fire propagation on a square grid.

## 2.2 Rules of Evolution

The CA model employs discrete states to characterize the evolution of each cell (i, j) at a given time step t. The table below outlines specific rules that determine how cells behave and transition within the model, significantly impacting the dynamics of fire spread.

**Table 1.** CA Fire Spread Model Transition Rules

| Rule | Current State | Next State | Description | Example |
|------|---------------|------------|-------------|---------|
| Rule 1 | 1 (TREE) | 1 (TREE) | Areas that contain unignited vegetation or are devoid of burning fuel remain unchanged. | If a cell contains unignited vegetation, it remains in the same state in the next time step. |
| Rule 2 | 1 (TREE) | 2 (BURNING) | Probability of tree ignition determined by a sigmoid function considering burning neighbors, vegetation density, and other factors. | If burning neighbors exceed a threshold, 'TREE' transitions. Probability of 'TREE' catching fire can be dynamically adjusted. |
| Rule 3 | 2 (BURNING) | 3 (BURNING DURATION) | Actively burning cells transition to the "burning duration" state in the subsequent time step. | If a cell is actively burning in the current time step, it turns into the "burning duration" state on the likelihood of neighboring cells (Moore neighborhood) in the next time step. |
| Rule 4 | 3 (BURNING DURATION) | 3 (BURNING DURATION) Varies based on pb | Probability of neighbouring cells transitioning to the "burning" state. | If a cell is actively burning, there's a probability (pb) that one or more of its neighbouring cells (i ± 1, j ± 1) will also be on fire in the next time step. |
| Rule 5 | 3 (BURNING DURATION) | (EMPTY) | Cells that have experienced complete combustion remain unchanged. | If a cell has already burned, it remains in the same state in the next time step. |

The local rule $f$ can be defined as:

$$f: S_{(i,j)}^{t} \times S_{N(i,j)}^{t} \to S_{(i,j)}^{t+1}$$

- f is a function that takes two arguments:
  - $S_{(i,j)}^{t}$: The present state of the central cell at position (i, j) at time t.
  - $S_{N(i,j)}^{t}$: The states of the neighbouring cells in the neighbourhood of cell (i, j) at time t.

The function f determines the future state $S_{(i,j)}^{t+1}$ of the central cell at position (i, j) at time t+1 based on its present state and the states of its neighbouring cells [15].

### 2.3 Factors Influencing Fire Spread

Our methodology comprehensively considers variables that profoundly impact the shape and rate of wildfire propagation:

1. **Type and Density of Vegetation:** Vegetation's type (e.g., agricultural, shrubs, pine trees) and density (sparse to dense) are very important. We classify these into discrete categories, assigning probabilities ($p_{veg}$ and $p_{den}$) that influence fire spread. We also used a mix of empty cells and tree cells to control vegetation density, thereby altering fuel load and fire propagation potential.

2. **Wind Speed and Direction:** Wind's role in fire dynamics is critical. We utilize an adaptable empirical equation to calculate fire spread probability ($p_w$) by incorporating wind velocity and direction. This approach accommodates various wind directions, enhancing model realism.

3. **Spotting Effect:** Fire spotting, where embers ignite new fires ahead of the main fire, is incorporated indirectly through the combined influence of wind and ignition sources.

## 3 CA-based simulation

The proposed model harnesses the flexibility of probabilistic Cellular Automaton (CA) techniques to dynamically simulate the intricate dynamics of forest fires across diverse landscapes. In this paradigm, the landscape is effectively partitioned into cells, forming a matrix where each cell possesses distinct attributes. This 2D grid serves as the spatial foundation upon which the probabilistic Cellular Automaton (CA) operates.

The essence of the probabilistic CA model lies in its ability to incorporate stochastic elements that mimic real-world fire dynamics. Each cell within the landscape possesses an inherent state, such as 'EMPTY', 'TREE', 'BURNING', 'BURNING DURATION', 'WATER', or 'CITY'. The transitions between these states are governed by a set of probabilistic rules, driven by factors such as the presence of neighbouring burning cells, the likelihood of lightning strikes $p_{lightning}$ and the impact of wind effects $Wind_{Speed}$ and $Wind_{Direction}$.

This probabilistic approach introduces a layer of uncertainty into the simulation, mirroring the unpredictable nature of fire behaviour in actual forest environments. The transition probabilities are calculated based on the state of neighbouring cells and external influencing factors. For instance, if the sum of burning neighbouring cells exceeds a predefined threshold (threshold), a 'TREE' cell is ignited, transitioning to the 'BURNING' state. These probabilities can be dynamically adjusted to reflect changes in factors such as fuel density, moisture content, wind speed, topographical features, and even localized climate patterns.

The probabilistic rules' variability is a foundational strength of the model. For instance, the threshold required for ignition of a 'TREE' cell can be dynamically modified to account for varying levels of available fuel, fostering accurate depictions of real-world fire susceptibility. Likewise, the likelihood of lightning strikes ($p_{lightning}$) can be tuned to accommodate differing weather conditions and historical lightning activity in the region.

Incorporating wind effects is another illustrative example of adaptability. The wind's influence on fire spread, quantified by $Wind_{Speed}$ and $Wind_{Direction}$ can be fine-tuned to mirror localized wind patterns, resulting in more precise predictions of fire behaviour. This approach enables the model to capture the intricate interplay between variables, further enhancing its predictive capabilities.

Moreover, the transition probabilities between states are amenable to customization. The model can be configured to account for different burning durations, with probabilities adjusted to represent a variety of fire intensities. This flexibility extends to cell states such as 'BURNING DURATION', where the transition probability can mirror the actual likelihood of a fire sustaining itself over time.

This probabilistic CA model embraces the randomness of fire-related events, capturing the interplay of neighbouring cell states, probabilistic influences, and the evolving nature of fire spread. It works as an excellent tool for simulating complex wildfire scenarios and enhancing our understanding of fire behaviour within diverse landscapes. In essence, the spatial distribution of probabilities, encoded within the network topology, facilitates the realistic emulation of fire dynamics, advancing the predictive capabilities of fire management strategies.

### 3.1 Tree Ignition Probability

Trees have a probability of catching fire, which is determined by a sigmoid function. This function maps the number of burning neighbours to a probability value between 0 and 1. As the number of burning neighbours increases, the probability of a tree catching fire also increases. The *sigmoid function* is defined as:

$$P_{ignition}(burning_{neighbours}) = 1 \Big/ \left(1 + exp(-k * (burning_{neighbour} - threshold))\right)$$

Where:

- $P_{ignition}$ is the probability of a cell catching fire.

- $burning_{neighbours}$ is the number of burning neighbors of the cell.

- *threshold* is a parameter that determines when ignition is likely.

- *k* is a scaling parameter that controls the steepness of the sigmoid curve.

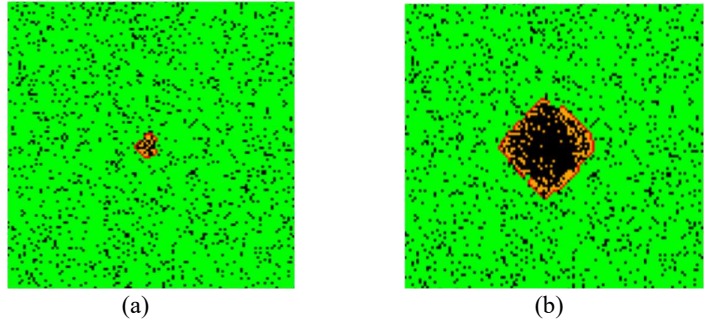

**Fig. 2.** (a) Initial configuration of the CA when it is ignited; (b) Configuration shows the spread of fire with 90% Vegetation Density

### 3.2 Wind Effects

The simulator takes into account the influence of wind on fire spread. Wind direction and speed affect the probability of fire spreading in a particular direction. The wind effects can be incorporated using a differential equation for fire spread:

$$dF/dt = k * Wind_{Speed} * Wind_{Direction} - m * F$$

Where:
- $F$ represents the fraction of cells that are on fire in the neighbourhood of the current cell.
- $dF/dt$ is the rate of change of the fraction of cells on fire.
- $Wind_{Speed}$ is the magnitude of the wind.
- $Wind_{Direction}$ is a vector representing the wind direction.
- $k$ is a constant that determines the influence of wind on fire spread.
- $m$ is a constant representing the fire extinguishing rate.

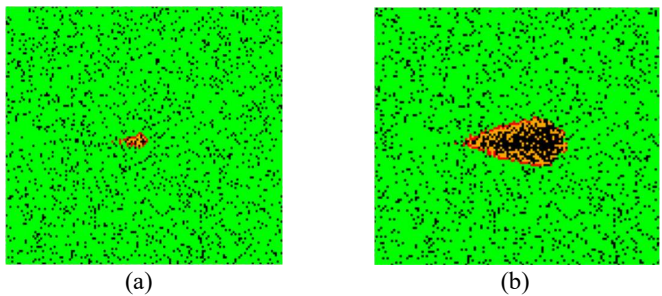

(a)          (b)

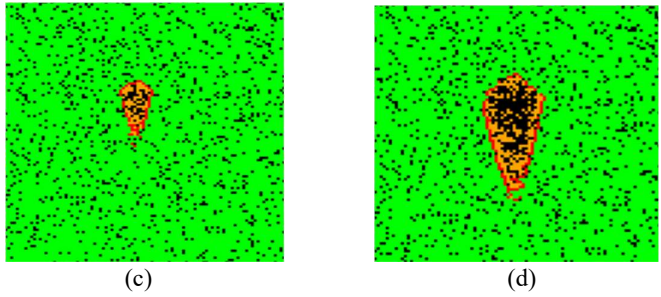

(c)                  (d)

**Fig. 3.** (a) Initial CA Configuration with Eastward Wind Direction; (b) Fire Spread Progression Eastwards; (c) Initial CA Configuration with Southward Wind Direction; (d) Fire Spread Progression Southwards

### 3.1 Calculation of Fire Spread Probability ($p_b$)

The probability of fire propagation $p_b$ is synthesized considering the above variables. We factor in a constant probability $p_o$ that a cell adjacent to a burning cell catches fire. This adjusted probability takes into account the combined effect of vegetation, density, wind, and terrain:

$$p_b = p_o(1 + p_{veg})(1 + p_{den})p_w$$

Where:

$p_o$ represents the baseline probability of fire propagation, derived from empirical data in the absence of wind effects, on flat terrain, and for a specific combination of vegetation density and type. This foundational probability serves as a reference point for further calculations. The components $p_{den}$, $p_{veg}$ and $p_w$ correspond respectively to the influences of vegetation density, vegetation type, and wind conditions (speed and direction) on the fire spread dynamics.

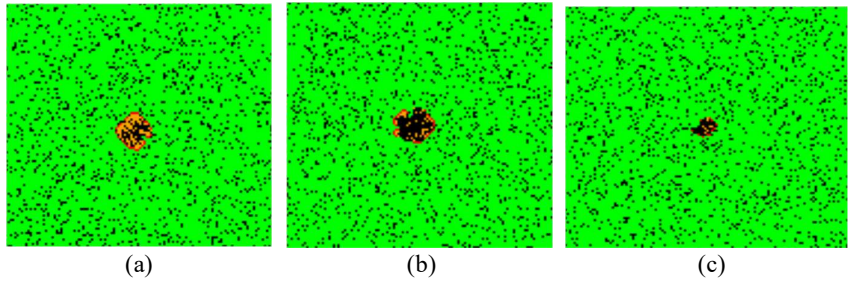

(a)                (b)                (c)

**Fig. 4.** (a) Configuration shows Fire Spread at Each Time Step with 90% Burning Probability; (b) Configuration shows Fire Spread at Each Time Step with 50% Burning Probability; (c) Configuration Shows Fire Spread at Each Time Step with 30% Burning Probability

## 4     Case Study: Wildfire on Spetses Island, 1990

Our proposed methodology was employed to predict the progression of a significant wildfire that ravaged Spetses Island on August 1, 1990. The fire originated from unknown sources within the island's forest, igniting near its central region. Propelled by moderate to strong north winds, the fire rapidly spread towards the south. Approximately 11 hours later, the fire was successfully extinguished after consuming nearly 6 km2, equivalent to about a third of the island's total area.

We selected this wildfire incident for two primary reasons. First, the incident's details, such as the extent of burnt area and time to extinguishment, were meticulously documented by local authorities. Second, the island's distinct topographical features, provided an ideal testing ground for evaluating the efficacy of our proposed Cellular Automaton (CA)-based model.

Initiating the application of our methodology involved generating matrices reflecting vegetation density and type. To achieve this, we created shape-files coding the island's vegetation type and density, drawing from pre-wildfire photomaps. Additionally, a shape-file delineating the burned area was generated. These digital datasets were then superimposed onto the grid to construct a vector data file. We adopted a square grid, to ensure precise representation of geographical data. From this vector data file, we derived matrices corresponding to vegetation density and type, alongside a vegetation density matrix and the aggregate burnt area.

Spetses Island hosts three vegetation types: Aleppo-pine trees, shrubs, and agricultural regions. The vegetation density was classified into three distinct categories: sparse, normal, and dense. Drawing from a combination of well-documented incident details and modern GIS tools, this case study offers a tangible platform to evaluate the capabilities of our CA-based model, as applied to real-world wildfire scenarios.

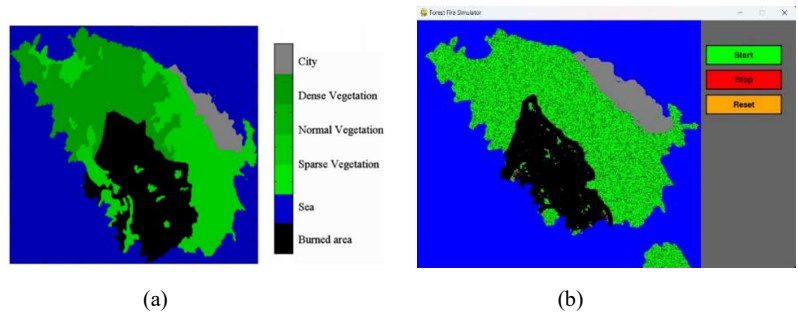

(a)                                            (b)

**Fig. 5.** (a) Actual Burned Area (Recreated from Photomap) [5]; (b) Predicted Burned Area from the Simulation.

## 5     Case Study of the 2021 Evia Island Wildfire, Greece

For our analysis we took as paradigm the case of the 2021 Greece wildfires, an illuminating case study emerged on Evia Island, Greece. Amidst a historic heatwave driving temperatures to an unprecedented 47.1 °C (116.8 °F), multiple wildfires ravaged the nation, inflicting immense damage. Against this backdrop, the largest wildfires tore through regions like Attica, Olympia, Messenia, and most devastatingly, northern Euboea. The second one suffered the most, with more than 50,000 hectares burned. Despite these difficulties, our method using Cellular Automaton (CA) was highly accurate for predicting wildfires.

The fires of 2021 scarred approximately 125,000 hectares of forest and arable land, echoing the 2007 Greek forest fires in magnitude. In this turbulent weather, a separate blaze struck Rhodes Island, prompting mass evacuations and disruptions. Amid this destruction, the 2021 Evia Island wildfire case study showed how effective our CA-based model can be.

By carefully combining data about incidents, modern GIS tools [9,10] and the island's detailed terrain, our model demonstrated it can accurately simulate real wildfires. As climate change makes wildfires more dangerous, CA-based methods show us a way to predict and manage wildfires ahead of time, helping us adapt to changing challenges.

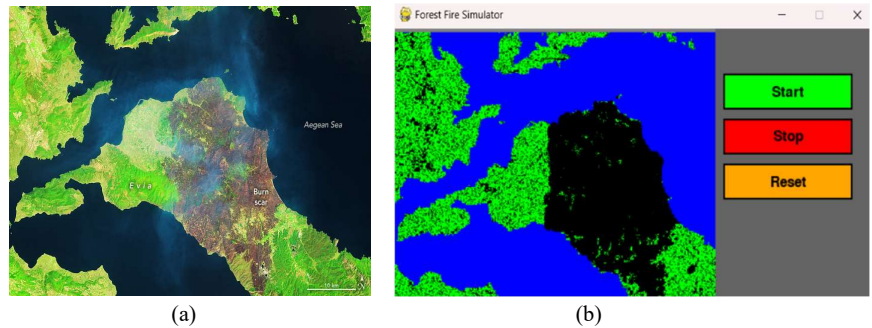

(a)             (b)

**Fig. 6.** (a) Actual Satellite Image Captured by NASA [16]; (b) Predicted Burned Area from the Simulation.

## 6     Conclusion

This paper has introduced an enhanced Cellular Automaton (CA) methodology designed to dynamically forecast wildfire spread. This approach factors in a multitude of influences on fire propagation, incorporating both meteorological and spatial geographical variables.

The practical application of our methodology entailed the simulation of a genuine wildfire that engulfed a renowned Greek island, destroying nearly half of its forested land.

The island's intricate terrain, featuring abrupt elevation changes and diverse vegetation characteristics, provided an ideal testing ground. To optimize the model's performance, a non-linear optimization technique was employed to fine-tune certain coefficients, minimizing the disparity between actual and predicted burnt areas. This optimization process involves running the simulation multiple times with different sets of parameter values, comparing the simulated results to real data, and iteratively adjusting the parameters to minimize the difference between the two. Encouragingly, the simulation outcomes closely aligned with real-world observations, affirming the potential of our methodology in effectively predicting fire spread across heterogeneous landscapes.

While the model has demonstrated promising results, its robustness will undergo further validation through application to diverse large-scale, real-world fire incidents. Ongoing efforts are directed towards improving the model's accuracy by incorporating various factors such as different types of vegetation, altitude, elevation, and humidity. Also recognizing the significant role of near-surface airflow patterns in fire front evolution, further improvements in the model is anticipated to enhance prediction accuracy.

In light of the accomplished work, future works should integrate additional parameters such as humidity, terrain features, and vegetation density. By expanding the model's scope, we aim to achieve higher simulation accuracy and extend its applicability to a broader range of wildfire scenarios. This approach holds the promise of improving wildfire prediction and management strategies, offering proactive solutions to mitigate the devastation caused by such events.